\documentclass[twocolumn,aps,prb,superscriptaddress,amstex,amsmath]{revtex4-2}
\usepackage{xcolor}
\usepackage{amsmath}
\usepackage{graphicx}

\makeatletter

\providecommand{\tabularnewline}{\\}

\usepackage{dcolumn}
\usepackage{bm}
\usepackage{gensymb}

\makeatother

\begin{document}
\title{Campbell penetration depth in low carrier density superconductor YPtBi}
\author{Hyunsoo Kim}
\email{hyunsoo.kim@ttu.edu}

\affiliation{Ames Laboratory, US Department of Energy, Ames 50011, IA, USA}
\affiliation{Department of Physics \& Astronomy, Iowa State University, Ames 50011, USA}
\affiliation{Maryland Quantum Materials Center, Department of Physics, University of Maryland, College Park, MD 20742, USA}
\affiliation{Department of Physics and Astronomy, Texas Tech University, Lubbock, Texas 79410, USA}
\author{Makariy A. Tanatar}
\affiliation{Ames Laboratory, US Department of Energy, Ames 50011, IA, USA}
\affiliation{Department of Physics \& Astronomy, Iowa State University, Ames 50011, USA}
\author{Halyna Hodovanets}
\affiliation{Maryland Quantum Materials Center, Department of Physics, University of Maryland, College Park, MD 20742, USA}
\affiliation{Department of Physics and Astronomy, Texas Tech University, Lubbock, Texas 79410, USA}
\author{Kefeng Wang}
\affiliation{Maryland Quantum Materials Center, Department of Physics, University of Maryland, College Park, MD 20742, USA}
\author{Johnpierre Paglione}
\affiliation{Maryland Quantum Materials Center, Department of Physics, University of Maryland, College Park, MD 20742, USA}
\author{Ruslan Prozorov}
\email{prozorov@ameslab.gov}

\affiliation{Ames Laboratory, US Department of Energy, Ames 50011, IA, USA}
\affiliation{Department of Physics \& Astronomy, Iowa State University, Ames 50011, USA}
\date{\today}


\begin{abstract}
Magnetic penetration depth, $\lambda_{m}$, was measured as a function of temperature and magnetic field in single crystals of low carrier density superconductor YPtBi by using a tunnel-diode oscillator technique.
Measurements in zero DC magnetic field yield London penetration depth, $\lambda_{L}\left(T\right)$, but in the applied field the signal includes the Campbell penetration depth, $\lambda_{C}\left(T\right)$, which is the characteristic length of the attenuation of small excitation field, $H_{ac}$, into the Abrikosov vortex lattice due to its elasticity. 
Whereas the magnetic field dependent $\lambda_C$ exhibit $\lambda_{C}\sim B^{p}$ with $p=1/2$ in most of the conventional and unconventional superconductors, we found that $p\approx 0.23\ll1/2$ in YPtBi due to rapid suppression of the pinning strength. 
From the measured $\lambda_{C}(T,H)$, the critical current density is $j_{c}\approx40\,\mathrm{A}/\mathrm{cm^{2}}$ at 75 mK. 
This is orders of magnitude lower than that of conventional superconductors of comparable $T_{c}$.
Since the pinning centers (lattice defects) and vortex structure are not expected to be much different in YPtBi, this observation is direct evidence of the low density of the Cooper pairs because $j_{c}\propto n_s$. 
\end{abstract}

\maketitle

\section{Introduction}

The superconducting phase transition temperature $T_{c}$ of the Bardeen-Cooper-Schrieffer (BCS) superconductors is typically of the order of $\sim10^{-4}T_{F}$, where $T_{F}\sim 10^4$ - 10$^5$ K is the Fermi temperature \cite{Bardeen1957,Tinkham2004}. 
This follows from the small value of the energy gap in the density of states, $\Delta\left(0\right)\approx1.76k_{B}T_{c}\approx2\hbar\omega_{D}\exp\left(-1/VN_0\right)$.
For example, for $T_{c}=10$ K, $\Delta\left(0\right)\approx1.5$ meV$\ll E_{F}$ $\sim 1$ - 10 eV. 
Here $\omega_{D}$ is the Debye frequency, $V$ is the attractive pairing potential, $N_0$ is the density of states (DOS) at the Fermi energy $E_{F}$.
$N_0$ has an exponential role in determining the $T_{c}$ which is often estimated from the semi-phenomenological McMillan equation \cite{McMillan1968,Allen1975}: 
\begin{equation}
T_{c}\approx\frac{2\hbar\omega_{D}}{1.76k_{B}}\exp\left[\frac{-1.04(1+\lambda)}{\lambda-\mu^{*}\left(1+0.62\lambda\right)}\right] 
\end{equation}
where $\lambda$ is the electron--phonon coupling parameter and $\mu^{*}$ is the screened Coulomb interaction constant. 
The weak-coupling BCS formula can be recovered by replacing $(\lambda-\mu^{*})$ with the product $VN_0$ in the limit of small $\lambda\ll1$. 
This relation was successfully applied for many intermetallic compounds with a typical carrier density $n\sim10^{22}$ cm$^{-3}$ \cite{DYNES1972}. 

However, discovery of superconductivity in materials with a poor metallic normal state with $n\sim 10^{17}$ - $10^{19}$ cm$^{-3}$ challenged the conventional approach.
Such low $n$ superconductors include elemental bismuth \cite{Prakash2017}, SrTiO$_{3}$ \cite{Lin2013}, and half-Heusler compound RTBi (R=rare earth, T=Pd or Pt) \cite{Goll2008,Butch2011,Tafti2013,Nakajima2015}.
The observed $T_{c}$s (and often critical fields) in these materials are by orders of magnitude higher than the expected $T_c$ from the McMillan formula \cite{Meinert2016}. These low $n$ superconductors have naturally much higher values of the ratio $T_{c}/T_{F}$, pushing them closer to the Bose-Einstein condensation (BEC) regime in which the spatial range of attractive interaction in the Cooper pair, the superconducting coherence length, $\xi$, becomes comparable to the characteristic length associated with the Fermi momentum, $\xi\sim\hbar/p_{F}$. 
In YPtBi the $k_{F}\approx 0.4$ nm$^{-1}$ whereas in SrPd$_{2}$Ge$_{2}$ with a similar $T_c$ it is $k_{F}\approx 10$ nm$^{-1}$. 
There are materials believed to be in the BCS-BEC crossover regime, notably FeSe$_{1-x}$S$_{x}$ \cite{Kasahara2014}.

Determination of the Cooper-pair density $n_{s}$ is required to confirm the unusual low $n$ nature of superconductivity in the material of interest.
Traditionally, the normal state electronic concentration $n$ is used to estimate $n_{s}$ by using a simple relation $n_{s}=n/2$. 
While the relation usually holds in the normal metals, accurate measurements of $n$ in the normal state with $E_{F}\ll1$ eV can be challenging due to strong temperature dependence of $n$ at low temperatures, anomalous Hall effect, and the presence of the surface states.
Here we probe $n_{s}$ directly in the superconducting state of YPtBi by determining the theoretical critical current density $j_{s}$, the quantity directly proportional to $n_{s}$.

The half-Heusler compound YPtBi is a topological semimetal with $n\sim10^{18}$ cm$^{-3}$ at low temperatures \cite{Lin2010,Butch2011}. 
Its superconductivity is attracting a considerable attention because $T_{c}$ is about fourfold higher than that of doped SrTiO$_{3}$ with a similar $n\sim10^{18}$ cm$^{-3}$, and it was suggested that its superconductivity arises from the $j=3/2$ Fermi surface \cite{Kim2018}. 
The possible superconducting states include unprecedented spin-quintet and septet states \cite{Brydon2016,Kim2018}.
The topological normal state is driven by strong spin-orbit coupling that inverts the $s$-orbital derived $\Gamma_{6}$ band and the $p$-orbital derived $\Gamma_{8}$ band \cite{Kim2018}. The chemical potential lies about 35 meV below a quadratically touching point of $\Gamma_{8}$ bands \cite{Butch2011,Kim2018} due to naturally occurring crystal imperfection \cite{Yu2017}.

Recent experimental results support unconventional superconductivity in YPtBi.
$T_{c}$ can be enhanced by physical pressure with an initial linear rate of 0.044 K/GPa \cite{Bay2012}. 
The upper critical field at zero temperature is $\mu_{0}H_{c2}(0)=1.5$ T \cite{Butch2011} that is higher than the Pauli limiting field 1.4 T \cite{Butch2011} for a weak-coupling spin-1/2 singlet superconductor. The temperature dependence of $H_{c2}(T)$ is practically linear over almost entire superconducting temperature-range, quite different from conventional parabolic behavior \cite{Butch2011}.
A muon spin rotation study determined $\lambda_{L}(0)=1.6$ $\mu$m \cite{Bay2014} which is an order of magnitude greater than that of strong type II superconductor CeCoIn$_5$ where $\lambda(0)\approx 0.26$ $\mu$m \cite{Chia2003}. Coherence length at zero temperature is $\xi(0)=\sqrt{\phi_{0}/2\pi H_{c2}\left(0\right)}\approx15\:\mathrm{nm}$. The Ginzburg-Landau parameter is $\kappa=\lambda_{L}(0)/\xi(0)\approx10^{2}\gg1/\sqrt{2}$, placing it in the strong type-II regime of superconductivity.

In the mixed state of a type II superconductor, the small-amplitude AC magnetic penetration depth is governed by the elastic properties of the Abrikosov vortex lattice in the linear response regime. 
This means that the amplitude of the AC field excitation, $H_{ac}$, is not large enough to displace the vortex out of the pinning potential well, and it only perturbs the vortex position within the validity of Hooke's law. 
In this case, the penetration depth is described by the Campbell penetration depth $\lambda_{C}$ that determines the attenuation range of the AC perturbation from the sample surface to the interior, $B_{ac}\left(x\right)\propto\mu H_{ac}e^{-x/\lambda_{C}}$ in a semi-infinite superconductor \cite{Campbell1969,Campbell1971,Brandt1995,Prozorov2003e,Willa2015prb,Willa2015,Willa2016}.
Here $\mu$ is magnetic permeability, and $x$ is the distance from the surface. Since $\lambda_{C}$ is not commonly measured due to amplitude/sensitivity limitations of the conventional AC techniques, we provide a simple derivation of $\lambda_{C}$ in the Appendix for completeness. The important advantage of employing $\lambda_C$ is that it gives access to the shielding current density via a relation $\lambda_{C}^{2}=H_{0}r_{p}/j_{c}$. Here $r_{p}$ is the radius of the pinning potential, and $H_{0}$ is the applied external DC magnetic field (see Appendix for details).
Importantly, the critical current density is estimated at the frequency of the measurement, and the rf regime gives access to almost unrelaxed values. The initial vortex relaxation is exponential, and hence the conventional techniques estimate relaxed values far from the true $j_{c}$ \cite{Blatter1994,Burlachkov1998,Matsushita2019}.


While analysis of the relaxed shielding current is complicated because of inclusion of the magnetic relaxation parameters, the unrelaxed critical current, $j=H_{0}r_{p}/\lambda_{C}^{2}$ offers direct access to the superfluid density $n_s\propto j_c$ \cite{Bardeen1958,Gorkov1959,Tinkham2004}.
In this work, we use a tunnel diode oscillator (TDO) technique to measure $\lambda_{C}(T,H)$ and determine $j_{c}(T,H)$ in YPtBi.
The determined $j_{c}$ is orders of magnitude smaller than that of well-known superconductors with the typical carrier density, and its rapid suppression by the magnetic field provides valuable insight into the fascinating nature of superconductivity at the low carrier density regime exemplified by YPtBi.

\section{Experimental}

\label{sec:exp}

YPtBi single crystals were grown out of molten Bi with starting composition Y:Pt:Bi = 1:1:20 (atomic ratio).\cite{Canfield1991,Canfield1992,Butch2011,Kim2018} The starting materials Y ingot (99.5\%), Pt powder (99.95\%), and Bi chunk (99.999\%) were put into an alumina crucible, and the crucible was sealed inside an evacuated quartz ampule. The ampule was heated slowly to 1150\degree C, kept for 10 hours, and then cooled down to 500\degree C at a 3\degree C/hour rate, where the excess of molten Bi was decanted by centrifugation.

The variation of the radio-frequency (rf) magnetic penetration depth $\Delta\lambda_{m}$ was measured in a dilution refrigerator by using a tunnel diode oscillator (TDO) technique  \cite{Kim2018rsi} (for review, see Ref. \cite{Prozorov2006,Prozorov2011}).

The sample with dimensions (0.29$\times$0.69$\times$0.24) mm$^{3}$ positioned with the shortest direction along $H_{ac}$ was mounted on a sapphire rod and inserted into a 2 mm inner diameter copper coil that (when empty) produces rf excitation field with amplitude $H_{ac}\approx20\:\mathrm{mOe}$ and frequency of $f_{0}\approx22\:\mathrm{MHz}$. The shift of the resonant frequency (in cgs units), $\Delta f(T)=-G4\pi\chi(T)$, where $\chi(T)$ is the differential magnetic susceptibility, $G\approx f_{0}V_{s}/2V_{c}(1-N)$ is a constant, $N$ is the effective demagnetization factor, $V_{s}$ is the sample volume and $V_{c}$ is the coil volume \cite{Prozorov2006}. The constant $G$ was determined from the full frequency change by physically pulling the sample out of the coil. With the characteristic sample size, $R$, $4\pi\chi=(\lambda_{m}/R)\tanh(R/\lambda_{m})-1$, from which $\Delta\lambda_{m}$ can be obtained \cite{Prozorov2006}.

\section{Results}

\begin{figure}
\includegraphics[width=0.95\linewidth]{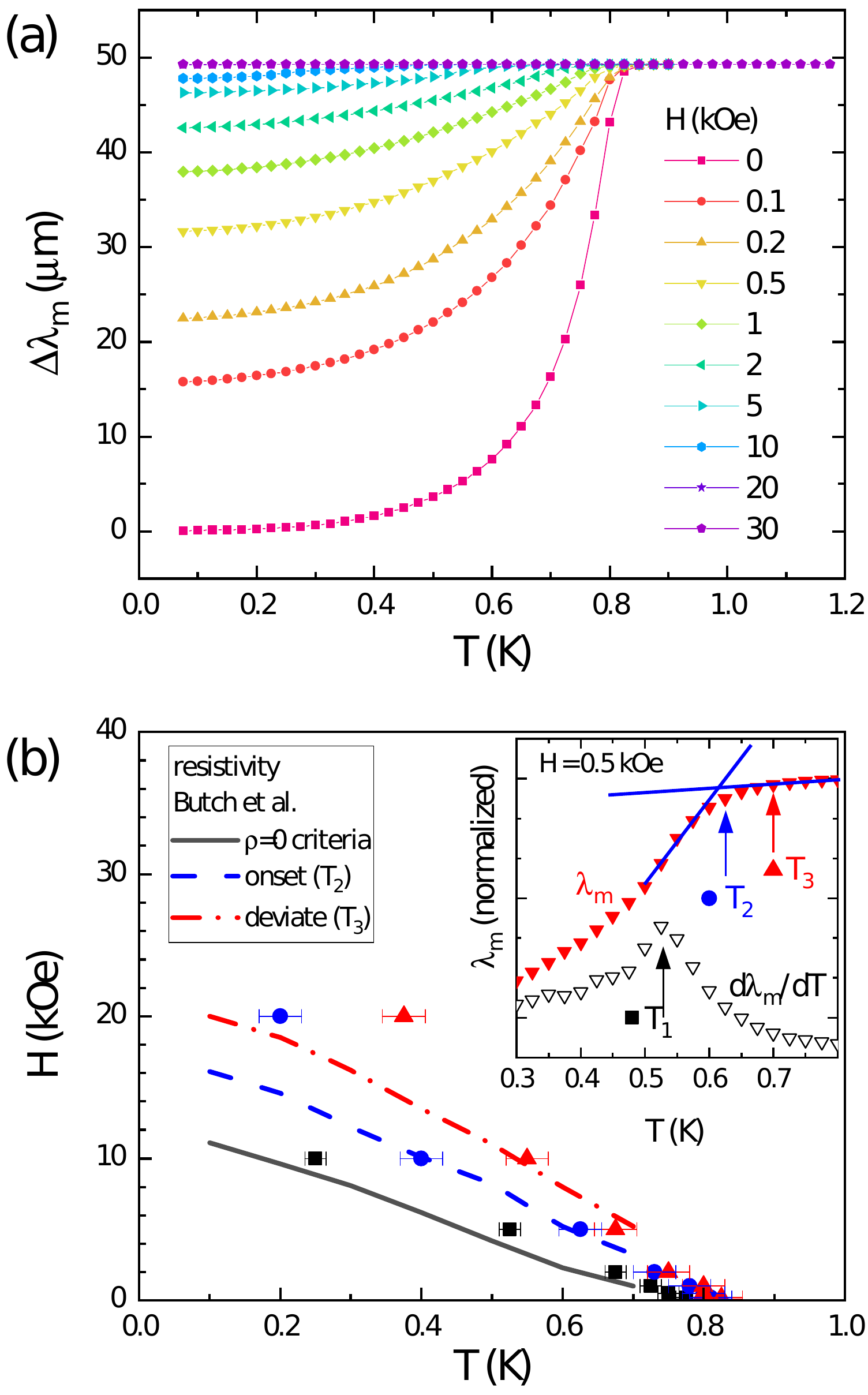}\caption{\label{fig1} (a) Temperature variation of the radio-frequency magnetic penetration depth, $\Delta\lambda_{m}(T)$, in various applied DC magnetic fields $H_{dc}$. (b) The $H$-$T$ phase diagram constructed from $\Delta\lambda_{m}(T,H)$ using characteristic temperatures, $T_{1}$ (maximum of $d\lambda_{m}/dT$, black squares), $T_{2}$ (crossing point of linear extrapolations, blue circles), and $T_{3}$ (onset of deviation, red up-triangles) as shown in the inset. The lines in the main panel of (b) show for reference the $H$-$T$ phase diagram as determined from the field-dependent electrical resistivity by zero resistivity (black), crossing point of linear extrapolations (blue dashes) and onset of deviation (red dash-dot) criteria \cite{Butch2011}.}
\end{figure}

Figure \ref{fig1}(a) shows temperature variation of the rf magnetic penetration depth $\lambda_{m}(T)$ in a single crystal of YPtBi in various applied DC magnetic fields $H_{dc}$ from 0 to 30 kOe (bottom to top). 
For $H_{dc}=0$, measured $\Delta\lambda_{m}(T)$ is the zero field limiting London penetration depth $\Delta\lambda_{L}(T)$ which exhibits a sharp superconducting phase transition at $T\approx0.8$ K.
We found $\Delta\lambda_{L}(T)=AT^{\alpha}$ where $A=1.98\:\mathrm{\mu m}/\mathrm{K}^{\alpha}$ and $\alpha=1.2$ \cite{Kim2018}. The observed exponent $\alpha$ is consistent with the presence of line-nodes in the superconducting gap and moderate impurity scattering. The large pre-factor, $A\propto\lambda_{L}(0)/\Delta(0)$ is compatible with a low carrier density superconductor within London theory $\lambda_{L}(0)=(mc^{2}/4\pi ne^{2})^{1/2}$. For comparison, $A=4$ - 15 \AA/K is observed in $d$-wave line-nodal high-temperature cuprate superconductors \cite{Prozorov2000} and 190 - 370 \AA/K in CeCoIn$_5$ \cite{Ormeno2002,Chia2003,Kim2015}.

Furthermore, YPtBi exhibits very pronounced field dependence of $\lambda_{m}$. 
It is notable even at the lowest $H_{dc}=100$ Oe which is 0.007$H_{c2}(0)$. Here we use $H_{c2}(0)=15$ kOe taken from Ref. \cite{Butch2011}. 
By measuring $\lambda_{m}(H,T)$ as a function of temperature in different applied fields we constructed the magnetic field-temperature $H$-$T$ phase diagram of YPtBi. 
Due to the broadness of the superconducting transition, we used three different criteria for the determination of $T_{c}$, as illustrated in the inset of Fig. \ref{fig1}(b). $T_{1}$ was determined at the sharp maximum of $d\lambda_{m}/dT$ (black squares). 
$T_{2}$ was determined at the intersection of the lines through the data in the superconducting state and the normal states (blue circles). 
$T_{3}$ was determined at the onset of $\Delta\lambda_{m}(T)$ deviation from the normal state behavior (red up triangles). 
The phase diagram from rf magnetic penetration depth data is shown in the main panel of Fig.~\ref{fig1}(b). For reference, we show the diagram as determined from resistivity measurements by Butch \textit{et al.} \cite{Butch2011}, using zero resistivity (black solid line), crossing point of linear extrapolations (blue dashes) and onset of deviation (red dash-dot) criteria. 
The phase diagram as determined from maximum of $d\lambda_{m}/dT$ line is closely following the phase diagram as determined from resistivity measurements using $\rho=0$ \cite{Butch2011}.
However, we detected apparent diamagnetic response in YPtBi at notably higher fields than in resistivity measurements, suggesting persistence of superconductivity in some non-bulk form \cite{Kim2013}. A discrepancy between $H_{c2}$ as determined from bulk thermal conductivity and non-bulk resistivity measurements is a known problem \cite{Tanatar2007} and it is usually assigned to the superconducting layer surviving at the surface. Perhaps, it is related to the third critical field in superconductors predicted theoretically by Saint-James and Gennes \cite{Saint-James1963} when a thin superconducting layer is formed on the flat surface parallel to the field. Recently, a surface-sensitive tunneling experiment on YPtBi detected energy gap spectra at higher temperatures than $T_{c}\approx0.8$ K \cite{Baek2015}. A similar signature of the superconducting phase was reported in another half-Heusler compound LuPtBi, which was attributed to the presence of van Hove singularity near $E_{F}$ \cite{Banerjee2015} and surface pairing states \cite{Wang2018}. The shape of tunneling spectra in the superconducting state is inconsistent with an isotropic $s$-wave in both YPtBi and LuPtBi \cite{Baek2015,Banerjee2015}.

\begin{figure}
\includegraphics[width=0.9\linewidth]{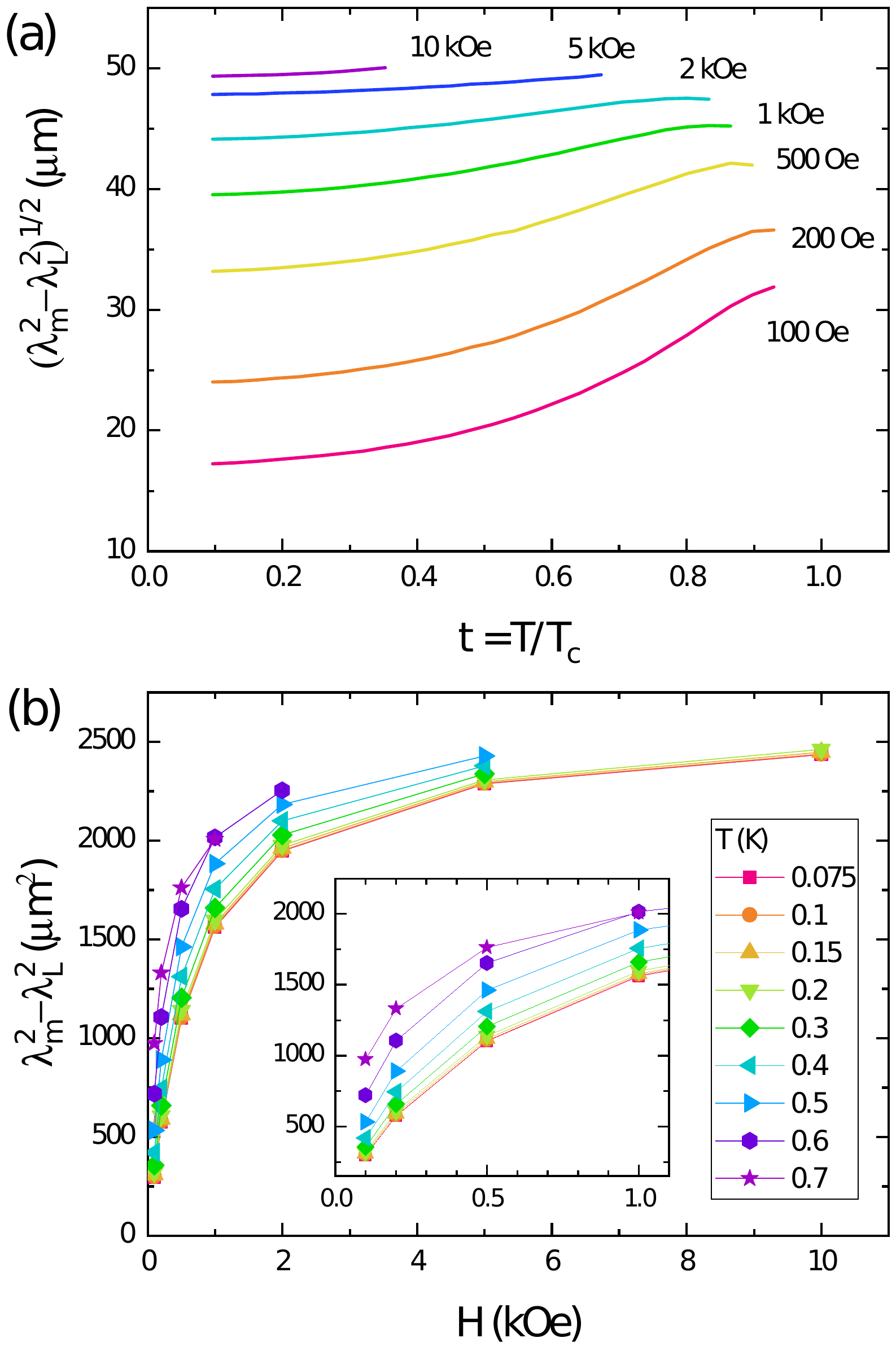}\caption{\label{fig2} (a) Temperature variation of the Campbell length $\lambda_{C}(T)=\sqrt{\lambda_{m}^{2}(T)-\lambda_{L}^{2}(T)}$ in DC magnetic fields $H_{dc}$ as indicated in the panel. (b) Isotherms of field variation of $\lambda_{C}^{2}(H)$ in YPtBi. Inset shows zoom of the low-field regime.}
\end{figure}

We focus now on the variation of $\lambda_{m}$ with finite $H_{dc}$ in the mixed state. 
The measured magnetic penetration depth satisfies the relation $\lambda_{m}^{2}=\lambda_{L}^{2}+\lambda_{C}^{2}$ in the approximation of a linear elastic response of a vortex lattice to a small amplitude AC perturbation $H_{ac}$ \cite{Campbell1969,Campbell1971,Brandt1995}.
The TDO technique is a perfect probe for this measurements because of small frequency, $f_{0}=22$ MHz, and small amplitude of the perturbation, $H_{ac}=20$ mOe. Since $\lambda_{L}(T)=\lambda_{L}(0)+\Delta\lambda_{L}(T)$ where $\lambda(0)=1.6~\mu$m \cite{Bay2014}, we can readily calculate $\lambda_{C}(T)$ in various $H_{dc}$. However, we took a conservative approach, assuming that this approximation is valid only at temperatures below $0.5T_{1}$ because $\lambda_{m}(T)$ becomes comparable to the size of the sample as temperature increases towards $T_{c}$.
The calculated $\lambda_{C}=\sqrt{\lambda_{m}^{2}-\lambda_{L}^{2}}$ at various $H_{dc}$ is shown in Fig. \ref{fig2}(a). In all measured $H_{dc}$, $\lambda_{C}(T)$ shows monotonic increase with temperature as the superconductor allows more penetration of the rf field with increasing temperature.

\begin{figure}[t]
\includegraphics[width=0.8\linewidth]{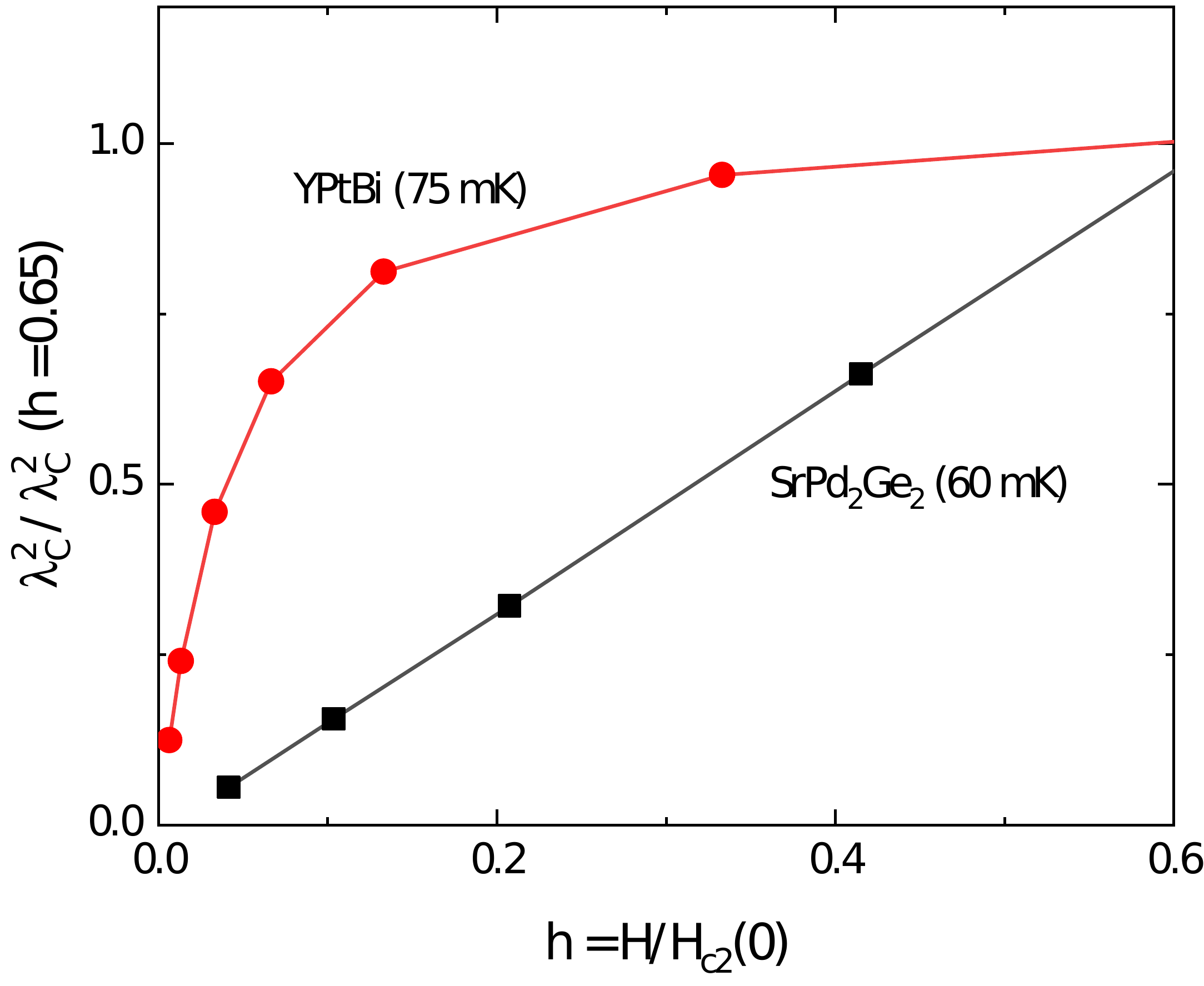}\caption{\label{fig3} Field-dependent Campbell penetration depth $\lambda_{C}^{2}(H)$ of YPtBi (red line with solid circles) shown in comparison with conventional metal/superconductor with similar $T_{c}$, SrPd$_{2}$Ge$_{2}$.
The data are normalized to the values determined at $0.65H_{c2}(0)$
for clarity.}
\end{figure}

Figure \ref{fig2}(b) shows the field-dependent $\lambda_{C}^{2}(H)$ at several temperatures. When critical current density does not vary much with field, we expect $\lambda_{C}^{2}(H)\sim H$ and it has been observed in most cases \cite{Prozorov2004,Prozorov2009,Kim2013}.
However, YPtBi exhibits significantly more curved $\lambda_{C}^{2}(H)$, indicating nearly logarithmic behavior at low fields.
This rapid rise of $\lambda_{m}(H)$ at low fields is unusual but may be explained considering very large values of $\lambda$ leading to strong intervortex interaction due to significant overlap already in low fields. In Figure \ref{fig3}, we compare this anomalous $\lambda_{C}(H)$ in YPtBi with SrPd$_{2}$Ge$_{2}$ which has a normal carrier density and exhibits $H$-linear behavior of $\lambda_{C}^{2}(H)$. 

\begin{figure}[t]
\includegraphics[width=1\linewidth]{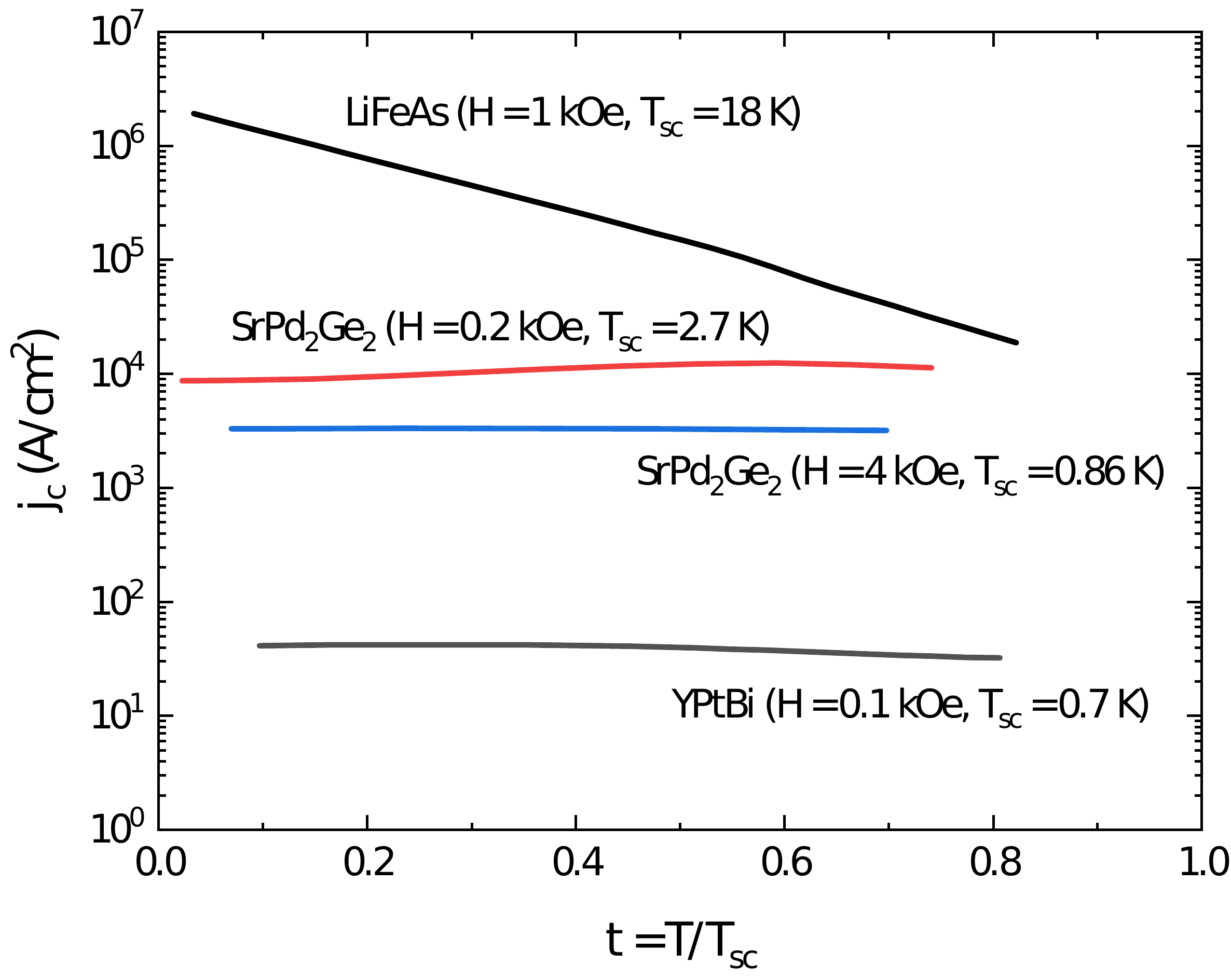}\caption{\label{fig4} Temperature variation of the theoretical current current density $j_{c}(T)$ in a selection of superconductors. We show the data in YPtBi in comparison with iron-based stoichiometric clean LiFeAs, and the low-temperature conventional superconductor SrPd$_{2}$Ge$_{2}$ measured in two different magnetic fields. $T_{sc}$ stands for the superconducting transition at a given magnetic field. For YPtBi, $T_{sc}=T_{1}$ (see Fig.~\ref{fig1}). }
\end{figure}

As noted above from the known $\lambda_{C}(T,H)$, one can evaluate critical current density $j_{c}$ via $j_{c}=H_{dc}r_{p}/\lambda_{C}^{2}$ (see appendix for details) where $r_{p}$ is a characteristic radius of the pinning potential, usually taken equal to the coherence length, $r_{p}(T)=\xi(T)$ \cite{Campbell1969,Campbell1971,Prozorov2004,Prozorov2009,Kim2013}.
Figure \ref{fig4} shows the calculated $j_{c}(T)$ in YPtBi, obtained from $\lambda_{C}(T)$ measurements taken in minimum applied $H_{dc}=100$ Oe. We compare $j_{c}(T)$ in YPtBi to the $j_{c}$'s determined in similar Campbell penetration depth measurements in some representative superconductors, LiFeAs \cite{Prommapan2011} and SrPd$_{2}$Ge$_{2}$ \cite{Kim2013}. The former is known as a two-band superconductor with full gaps \cite{Kim2011}, and the latter is a single-gap BCS superconductor \cite{Kim2013}. The compared $j_{c}(T,H)$ in these superconductors were obtained by using the same TDO technique. In particular, we used the same TDO setup for YPtBi and SrPd$_{2}$Ge$_{2}$.

In YPtBi, the highest $j_{c}(T)\approx40$ A/cm$^{2}$ is observed at $T\approx75$ mK, and $j_{c}(T)$ monotonically decreases with temperature. In LiFeAs, $j_{c}\approx1\times10^{6}$ A/cm$^{2}$ at the lowest temperature, and it monotonically decreases by two orders of magnitude upon warming. In SrPd$_{2}$Ge$_{2}$, $j_{c}\approx8.3\times10^{4}$ A/cm$^{2}$ at the lowest temperature with $H_{dc}=200$ Oe. However, its temperature variation is non-monotonic and exhibits a maximum at an intermediate temperature, which was attributed to a matching effect between temperature-dependent coherence length and relevant pinning length scale \cite{Kim2013}. At a higher $H_{dc}=4$ kOe, $j_{c}(T)$ recovers a monotonic decrease with increasing temperature.
Even when $T_{sc}(H)$ of SrPd$_{2}$Ge$_{2}$ was reduced to 0.86 K in $H_{dc}=$4 kOe which is close to $T_{sc}$ of YPtBi with 100 Oe, $j_{c}$ is still two orders of magnitude greater, and the difference gets even bigger at base temperature. It is also instructive to calculate the depairing current density at which Cooper pairs break apart reaching critical velocity, $4\pi c^{-1}j=\phi_{0}/\left(3\sqrt{3}\lambda_{L}^{2}\xi\right)\approx1\times10^{7}\:\mathrm{A/}\mathrm{cm^{2}}$, which is much larger than $j_{c}$ due to pinning, but two orders of magnitude less than in ``typical'' normal carrier density superconductors \cite{Blatter1994,Matsushita2019}. 

In Table \ref{tab1}, we compare normal state Hall constants $R_{H}$ reported for YPtBi ~\cite{Butch2011} and SrPd$_{2}$Ge$_{2}$~\cite{Sung2011}. In both compounds the Hall resistivity $\rho_{xy}(H)$ is field-linear, which enables $R_{H}$ definition from the slope of the curve and sample geometry. The reported $R_{H}$ values are $-1.6\times10^{-4}$ cm$^{3}$/C {[}\onlinecite{Sung2011}{]} and $+2.4$ cm$^{3}$/C {[}\onlinecite{Butch2011}{]} for SrPd$_{2}$Ge$_{2}$ and YPtBi, respectively. In the single band Drude model, the carrier density satisfies a simple relation, $R_{H}=1/ne$ where $e$ is the electron charge. The calculated carrier densities are electron-like $n_{e}=3.9\times10^{22}$ cm$^{-3}$ for SrPd$_{2}$Ge$_{2}$ and hole-like $n_{h}=2.6\times10^{18}$ cm$^{-3}$ for YPtBi. For reference we also present $j_{c}$ in both SrPd$_{2}$Ge$_{2}$ and YPtBi in Tab. \ref{tab1}.

\section{Discussion}

The carrier density $n$ is responsible for the observed theoretical current density because $j_{c}\propto n_{s}v_{s}$ where $v_{s}$ is the velocity of the supercurrent \cite{Gorkov1959,Bardeen1958,Tinkham2004}. Provided $n_{s}\approx n/2$, $j_{c}$ in YPtBi is two orders of magnitude smaller than that in SrPd$_{2}$Ge$_{2}$, while the normal state $n$ in YPtBi is smaller by four orders of magnitude. Consequently, $v_{s}$ in YPtBi is two orders of magnitude greater than that in SrPd$_{2}$Ge$_{2}$. In Ginzburg-Landau theory, $j_{c}$ is associated with the depairing velocity $v_{s}=\Delta(0)/\hbar k_{F}$, and since $\Delta(0)$ values in both superconductors are of the same order, the different $v_{s}$ is accounted by different $k_{F}$ values in these two compounds. In SrPd$_{2}$Ge$_{2}$, $k_{F}\approx10$ nm$^{-1}$ in free electron approximation, i.e., $k_{F}=(3\pi^{2}n)^{1/3}$ with $n=2.6\times10^{22}$ cm$^{-3}$ which is about two orders of magnitude greater than that in YPtBi, $k_{F}=0.37$ nm$^{-1}$. \cite{Kim2018}

\begin{table}[b]
\caption{\label{tab1}Hall constant $R_{H}$, $n$, and $j_{c}$ in YPtBi and
SrPd$_{2}$Ge$_{2}$. }

\begin{ruledtabular}
\begin{tabular}{c|ccc}
 & $R_{H}$ (cm$^{3}$/C)  & $n$ (cm$^{-3}$)  & $j_{c}$ (A/cm$^{2}$)\tabularnewline
\hline 
YPtBi  & $+2.4$ \cite{Butch2011} & $2.6\times10^{18}$  & 40\footnote{determined at 75 mK ($0.096T_c$) and 100 Oe ($0.007H_{c2}(0)$).} \tabularnewline
SrPd$_{2}$Ge$_{2}$  & $-1.6\times10^{-4}$ \cite{Sung2011} & $3.9\times10^{22}$  & 8300\footnote{determined at 60 mK ($0.022T_c$) and 200 Oe ($0.042H_{c2}(0)$).} \tabularnewline
\end{tabular}
\end{ruledtabular}

\end{table}

\begin{figure}
\includegraphics[width=1\linewidth]{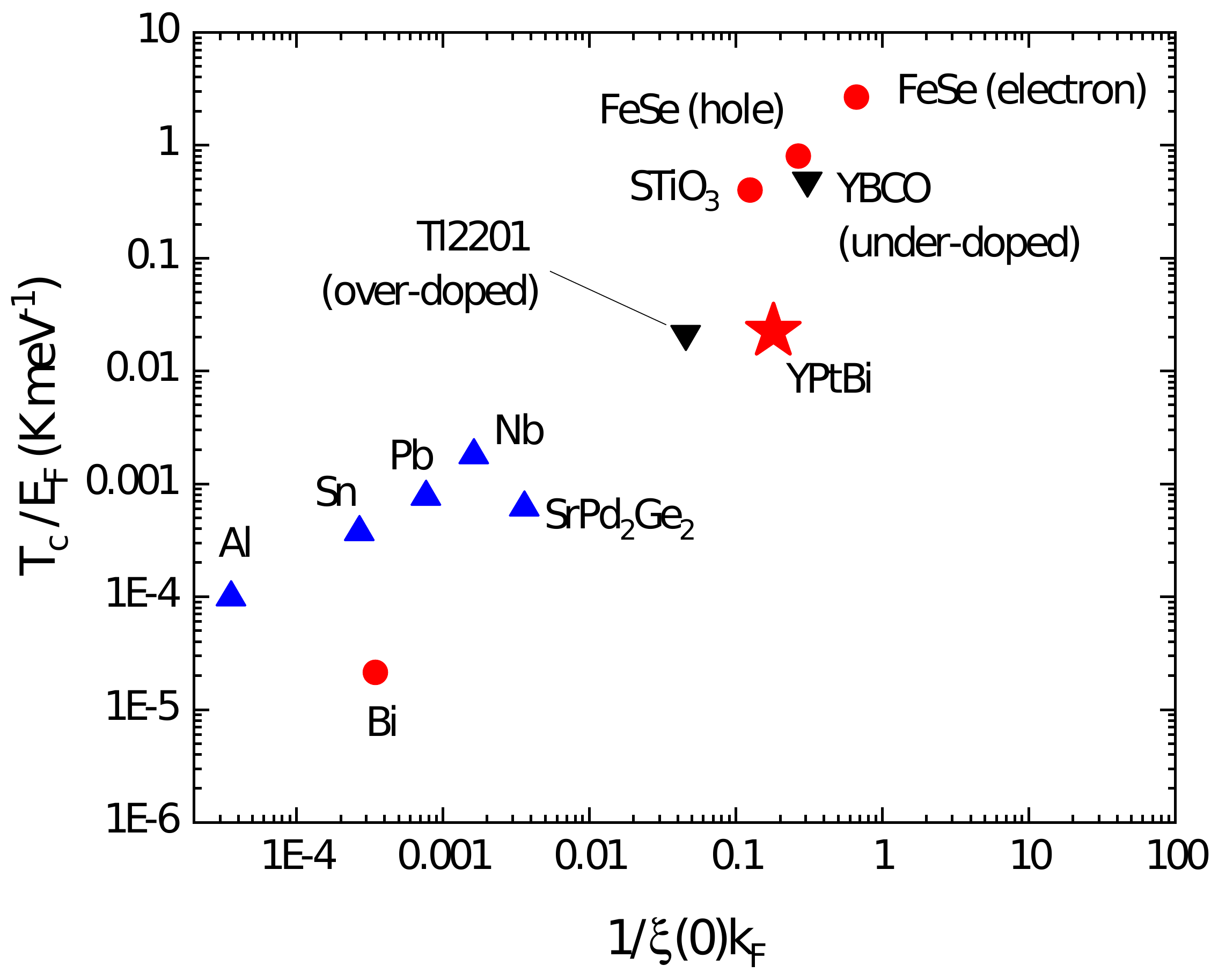}\caption{\label{fig5} Comparative BCS-BEC plot of $T_{c}/E_{F}$ vs. $1/k_{F}\xi(0)$ for low carrier density ( $n\ll10^{22}$ cm$^{-3}$, red circles), conventional metallic superconductors (blue up-triangles) and some representative cuprates, under-doped YBCO and overdoped Tl2201 (black down-triangles). Red star is the result for YPtBi. }
\end{figure}

With the confirmed low density of the Cooper pairs in the superconducting YPtBi, we can expect $E_{F}\sim n^{3/2}$ to be respectively low. In combination with the anomalously high $T_{c}$, not accounted for by the McMillan formula \cite{Meinert2016}, this leads to a large value of the $T_{c}/E_{F}$ ratio and a possibility of BCS-BEC crossover in YPtBi.

Superfluidity without Cooper pairing, i.e. BEC, can be achieved when the de Broglie wavelength of a particle becomes larger than the inter-particle distance, causing strong correlations. The possibility of BCS-BEC crossover was discussed in the low carrier density systems in which the effective size of the conduction electron is comparable to that of the Cooper pair. The mean spacing between conduction electrons, $r_{s}$, can be described with $k_{F}$ by the relation, $r_{s}\approx7.1/k_{F}$, for the parabolic band. In YPtBi, $r_{s}\approx19$ nm while the zero temperature coherence length is only $\xi(0)=15$ nm.

In the BCS-BEC crossover regime, the overlap between Cooper pairs is small or $\xi(0)\sim1/k_{F}$. Many superconductors have been tested for the conditions for BEC, and recently the multigap superconductor FeSe was found to be near the regime, in which $1/k_{F}\xi(0)\approx0.26$-$0.67$ \cite{Kasahara2014}. For YPtBi, $k_{F}\approx0.4$ nm$^{-1}$ \cite{Kim2018} and $\xi(0)=15$ nm \cite{Butch2011}, which makes $1/k_{F}\xi(0)\approx0.17$, and the ratio $T_{c}/T_{F}\approx2\times10^{-3}$ from $T_{c}\approx0.8$ K and $k_{B}T_{F}$ = 35 meV. Several low carrier density superconductors are compared to the well-known superconductors in the summary plot of $T_{c}/E_{F}$ vs. $1/k_{F}\xi(0)$ in Fig. \ref{fig5}. We find that both YPtBi (red star) and SrTiO$_{3}$ are relatively close to the crossover regime whereas Bi is well in the BCS limit. Tuning chemical potential in YPtBi and SrTiO$_{3}$ by gating or charge doping, particularly towards smaller $E_{F}$, would be uppermost interesting for understanding their pairing mechanism.

There has been much effort to elucidate the unconventional superconductivity in the low carrier density superconductors including YPtBi and SrTiO$_{3}$.
Recently, the unexpectedly high $T_{c}$ in YPtBi was explained by the electron-phonon pairing mechanism with polar optical phonon mode within the $j=3/2$ Luttinger-Kohn four-band model~\cite{Savary2017}.
In the similar low carrier density superconductors, the plasmonic~\cite{Ruhman2016} and nonadiabatic~\cite{Gorkov2016} superconducting mechanisms were proposed in SrTiO$_{3}$. 

The structure of the superconducting energy gap and the symmetry of pairing interaction are prerequisites for understanding the superconducting mechanism, but low $T_{c}$ in the low $n$ superconductors makes the experimental investigation difficult. The half-Heusler compounds RTBi (R=Y,La,Lu, T=Pt,Pd) exhibit relatively high superconducting transition temperatures $T_{c}\sim1$ K\cite{Butch2011,Tafti2013,Nakajima2015}, and a nodal superconducting gap was observed in YPtBi~\cite{Kim2018}.
Subsequently, various exotic pairing symmetries were proposed including nematic $d$-wave~\cite{Boettcher2018,Sim2019} and $j=3/2$ high-spin superconductivity~ \cite{Brydon2016,Yu2017,Kim2018,Venderbos2018,Wang2018,Roy2019}. In general, the high-spin superconductivity exhibits topological gap structures with the possibility of harboring the Majorana surface fluid~\cite{Wang2017,Wang2018}, which makes the low carrier density superconductor RTBi a promising platform for the fault-tolerant quantum devices.

\section{Summary}

We measured rf superconducting magnetic penetration depth in single-crystal YPtBi. 
The London penetration depth is consistent with the nodal superconductivity in YPtBi. 
In the finite DC magnetic fields, the measured Campbell penetration depth exhibits unusual sub-quadratic power-law behavior in the low field range.
From the variation of the Campbell penetration depth, we estimated the theoretical critical current density which is orders of magnitude smaller than that of the superconductors with a typical carrier density.
Therefore, we confirmed the low carrier density nature of superconductivity in YPtBi.

\section{acknowledgments}

We would like to thank V. G. Kogan for the useful discussion. Work in Ames was supported by the U.S. Department of Energy (DOE), Office of Science, Basic Energy Sciences, Materials Science and Engineering Division. Ames Laboratory is operated for the U.S. DOE by Iowa State University under contract DE-AC02-07CH11358. Research done at the University of Maryland was supported by the U.S. Department of Energy (DOE) Award No. DE-SC-0019154 (experimental investigations), and the Gordon and Betty Moore Foundation EPiQS Initiative through Grant No. GBMF4419 (materials synthesis).

\appendix

\section{Illustration of the concept of Campbell length} 
Here we provide simple arguments behind the concept of pinning and Campbell penetration depth. This physics has been discussed multiple times in the past 80 years and is textbook material. However, we felt that it is instructive and to write down the derivation with units and show step by step the flow. There is still a significant degree of confusion dealing with currents, fields, and flux in cgs and SI. There is also some confusion about the Campbell penetration depth written for a single vortex featuring single flux quanta $\phi_{0}$ vs. what should be with the magnetic induction, $B$, recognizing that this is a collective effect. This is because the critical current density is introduced via the single vortex pinning. Unfortunately, the original derivation by Campbell \cite{Campbell1969,Campbell1971} is too short and too schematic to our taste and we wanted to explain everything step by step.

\subsection{Single vortex pinning}

\begin{figure}
\includegraphics[width=1\linewidth]{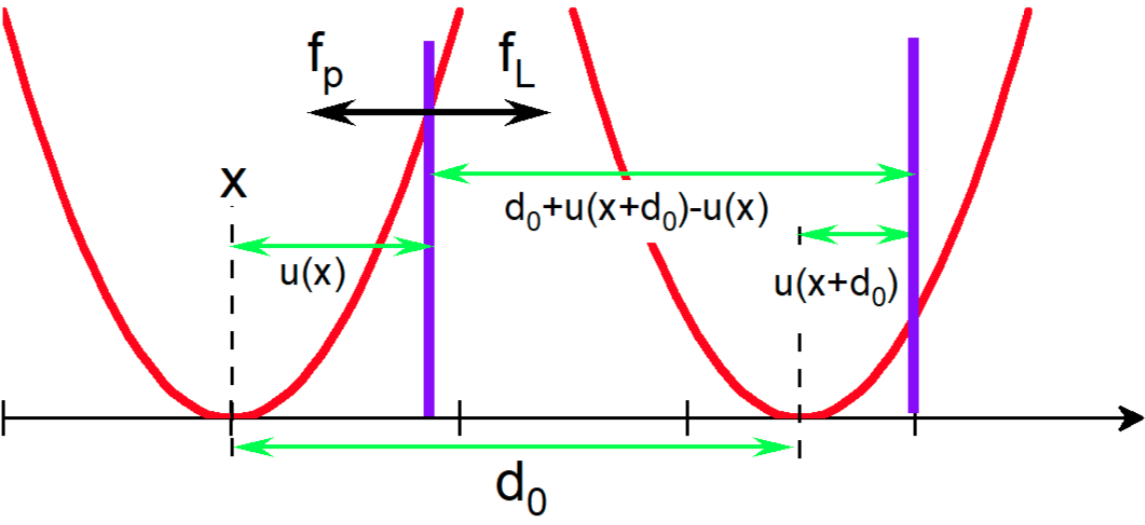}\caption{\label{fig6} Schematic for movement of vortices in the pinning potential. The violet vertical bars represent vortices, and the red parabolas are the pinning potential, $U(u)$. The vortices move along the horizontal line ($x$-axis), the electrical current flows into or out of the page ($y$-axis), and the external magnetic field is applied along the positive vertical axis ($z$-axis). The displacement $u$ stands for the horizontal position of each vortex from the equilibrium point. }
\end{figure}

\textit{Note, we use SI Units throughout Appendix section.}

Figure \ref{fig6} shows the schematics of the vortices and pinning potential model. The vortices move along the horizontal $x$-axis, the electrical current flows into or out of the page along the $y$-axis, and the external magnetic field is applied along the positive vertical direction of the $z$-axis. The displacement $u(x)$ is the deviation of a vortex from the center of its potential well. In equilibrium, $u(x)=0$ for all vortices and their distribution is constant. 

Assuming a vortex along the positive $z$-axis in a pinning potential $U(u)$ (real units energy-distance), then for a single vortex, the Lorentz force \textit{per unit length} is 
\begin{equation}
\mathbf{f}_{L}\left[\frac{\textmd{N}}{\textmd{m}}\right]=\mathbf{j}\left[\frac{\textmd{A}}{\textmd{m}^{2}}\right]\times\mathbf{\hat{z}}\mathbf{\phi}_{0}\left[\textmd{T m}^{2}\right]\label{fL}
\end{equation}
Here, ${\phi}_{0}=2.068\times10^{-15}$ Wb = T m$^{2}$.

The electric current density $\mathbf{j}$ along the positive $y$-direction would push the vortices to the positive $x$-direction with the magnitude, $f_L=j\phi_0$. It is usually assumed that the pinning potential is given by
\begin{equation}
U(u)=\frac{1}{2}\alpha u^{2}\left[\frac{\text{J}}{\text{m}}\right]
\end{equation}
where $\alpha$ is the so-called Labusch parameter: 
\begin{equation}
\alpha=\frac{d^{2}U(u)}{du^{2}}\left[\frac{\text{J}}{\text{m}^{3}}\right].
\end{equation}

The pinning force due to $U(u)$ is defined by,
\begin{equation}
f_{p}=-\frac{dU(u)}{du}=-\alpha u.
\end{equation}
In the presence of the electrical current, the two forces, $\mathbf{f}_{L}$ and $\mathbf{f}_{p}$, will act on a vortex in the opposite directions, i.e., $f_{L}+f_{p}=0$. In this case, the critical current density, $j_{c}$, is reached when the magnitudes of two forces become equal at a distance $u=r_{p}$ that is called the "radius of the pinning potential." In equilibrium, $j_{c}\phi_{0}=\alpha r_{p}$, and therefore $j_{c}$ can be expressed as:
\begin{equation}
j_{c}=\frac{\alpha r_{p}}{\phi_{0}}\left[\frac{\text{A}}{\text{m}^{2}}\right].\label{jcSV}
\end{equation}
The major contribution to pinning comes from the gain the free energy in the normal core volume of a vortex and, therefore, $r_{p}$ is usually assumed to be equal to the superconducting coherence length $\xi$. Of course, the pinning theory is much more complex, and the readers are referred to the excellent review articles, Ref.\cite{Blatter1994,Brandt1995}.

\subsection*{The Campbell penetration depth}

The Campbell length $\lambda_{C}$ is \textit{defined for a large number of vortices} since this is a wave-like perturbation in the vortex lattice treated as an elastic medium \cite{Blatter1994,Brandt1995}.  In words, $\lambda_{C}$ determines how far a small-amplitude ac perturbation on the superconductor edge propagates into the vortex lattice.) \cite{Campbell1969}  Let us assume a uniform distribution of vortices (e.g. after field-cooling, FC) and hence a uniform magnetic induction $B_{0}$. We apply a small ac field on the sample edge, i.e., $B=B_{0}+B_{ac}$, where $B_{ac} \ll B_{0}$. In equilibrium, the vortices are equally spaced by the distance $d_{0}$ found from the condition that each vortex carries a single flux quantum, $\phi_{0}$: 
\begin{equation}
d_{0}=\sqrt{\frac{\phi_{0}}{B_{0}}}=\frac{45.473}{\sqrt{B\left[\text{T}\right]}}\left[\text{nm}\right]\label{d0}
\end{equation}
Here we assumed a square vortex lattice instead of triangular for simplicity, which does not alter the results.

Consider a one-dimensional problem (semi-infinite superconductor positioned at $x\geq0$) with a magnetic field applied along the positive $z$-axis and electric current flowing along the positive $y$-axis. At a distance, $x$, from the edge, a row of vortices uniformly spaced along the $y$-axis is displaced by $u(x)$ from their equilibrium positions. The next row of vortices is displaced by the distance $d_{0}+u\left(x+d_{0}\right)-u(x)$ counted from the first row of vortices (see Fig. \ref{fig6}). Therefore, the distance between the vortices, $d(x)$, satisfies
\begin{equation}
d\left(x\right)=d_{0}\left[1+\frac{u\left(x+d_{0}\right)-u(x)}{d_{0}}\right]\approx d_{0}\left(1+\frac{du}{dx}\right)\label{d(x)}
\end{equation}
because $d_{0}$ is the smallest physical distance in the problem. Therefore, the magnetic induction $B$ at the location $x$ is given by 
\begin{equation}\label{Bofx}
B\left(x\right)=\frac{\phi_{0}}{dd_{0}}=\frac{\phi_{0}}{d_{0}^{2}\left(1+\frac{du}{dx}\right)}=B_{0}\left(1-\frac{du}{dx}\right)
\end{equation}
where we assume that $\frac{du}{dx}\ll 1,$ which can be easily checked with the final solution for $u(x)$. Note that if all vortices were displaced uniformly, $u=const$, then $B(x)$ remains unchanged.

This perturbation of $B(x)$ corresponds to the current density from the Maxwell equation, $\mu_{0}\mathbf{J}=\nabla\times\mathbf{B}$. Assuming $\mathbf{B}=B(x)\hat{z}$, 
\begin{equation}
\mu_{0}J_{y}=-\frac{\partial B\left(x\right)}{\partial x}=B_{0}\frac{d^{2}u}{dx^{2}}
\end{equation}
The Lorentz force on vortices per unit volume is
\begin{equation}
F_{L}=JB_{0}=\frac{B_{0}^{2}}{\mu_{0}}\frac{d^{2}u}{dx^{2}}
\end{equation}
which must be balanced by the pinning force. From the previous (single vortex) section, each vortex experiences pinning force per unit length, $f_{p}=-\alpha u$, and there are approximately $N=B_{0}/\phi_{0}$ vortices per unit area. The total pinning force \textit{per unit volume}, $F_{p}$, can be written in the form
\begin{equation}
F_{p}=Nf_{p}=-\frac{\alpha B_{0}}{\phi_{0}}u.
\end{equation}
The two forces, $F_{L}$ and $F_{p}$, balance each other in the steady-state, and the characteristic penetration depth is determined from the following relation,
\begin{equation}
F_{L}+F_{p}=\frac{B_{0}^{2}}{\mu_{0}}\frac{d^{2}u}{dx^{2}}-\frac{\alpha B_{0}}{\phi_{0}}u=0
\end{equation}
or
\begin{equation}\label{lambdaC}
\lambda_{C}^{2}\frac{d^{2}u}{dx^{2}}=u
\end{equation}
Here we introduced the Campbell length:
\begin{equation}
\lambda_{C}^{2}=\frac{\phi_{0}B_{0}}{\mu_{0}\alpha}\left[\frac{\text{T}^{2}\text{m}^{2}}{\frac{\text{H}}{\text{m}}\frac{\text{J}}{\text{m}^{3}}}=\text{m}^{2}\right]\label{lambdac2}
\end{equation}
Note that the radius of the pinning potential, $r_{p}$, does not explicitly enter here. This is true only for parabolic pinning potential within the validity of Hooke's law for vortex displacement. The non-parabolic potentials have also been considered and lead to a variety of interesting effects. \cite{Prozorov2003e,Gordon2013a,Willa2015,Willa2016} 

The Labusch constant $\alpha$ can be evaluated from the measured $\lambda_{C}$ by using (\ref{lambdac2}).
The solution of the equation (\ref{lambdaC}) for $u$ is
\begin{equation}
u\left(x\right)=u_{0}e^{-x/\lambda_{C}}
\end{equation}
Therefore the magnetic induction can be found by using (\ref{Bofx}) as follows:
\begin{equation}
B\left(x\right)=B_{0}\left(1-\frac{du}{dx}\right)=B_{0}\left(1+\frac{u_{0}}{\lambda_{C}}e^{-x/\lambda_{C}}\right)
\end{equation}
At the boundary, $x=0,$ and $B=B_{0}+B_{ac}$ where
\begin{align*}
B_{ac} & =B_{0}\frac{u_{0}}{\lambda_{C}}\\
u_{0} & =\lambda_{C}\frac{B_{ac}}{B_{0}}
\end{align*}
The displacement $u(x)$ in terms of $B_{ac}$ and $B_0$ can be written in the form
\begin{equation}
u\left(x\right)=\lambda_{C}\frac{B_{ac}}{B_{0}}e^{-x/\lambda_{C}}\label{u(x)}
\end{equation}
and we can express $B(x)$ as 
\begin{equation}
B\left(x\right)=B_{0}\left(1+\lambda_{C}\frac{B_{ac}}{B_{0}}\frac{u_{0}}{\lambda_{C}}e^{-x/\lambda_{C}}\right)=B_{0}+B_{ac}e^{-x/\lambda_{C}}\label{B(x)}
\end{equation}
which is expected from the boundary conditions.

Finally, we derive a practical expression for $J_{c}$ in terms of $\lambda_{C}$ that can be experimentally determined. Using (A18): 
\begin{align*}
\lambda_{C}^{2} & =\frac{\phi_{0}B_{0}}{\mu_{0}\alpha}=\frac{\phi_{0}H}{\alpha}\\
\frac{\phi_{0}}{\alpha} & =\lambda_{C}^{2}\frac{\mu_{0}}{B_{0}}
\end{align*}
Thus, $J_{c}$ is related to $\lambda_{C}$ as follows:
\begin{equation}
J_{c}=\frac{\alpha r_{p}}{\phi_{0}}=\frac{B_{0}r_{p}}{\mu_{0}\lambda_{C}^{2}}=\frac{H_{0}r_{p}}{\lambda_{C}^{2}}\label{Jc}
\end{equation}
We use the relation (\ref{Jc}) to calculate the critical current density from the measured Campbell penetration depth in the main text.

%

\end{document}